%% file: root.tex
\documentclass{ifacconf}

\usepackage{graphicx} 
\usepackage{epsfig}   
\usepackage{mathptmx} 
\usepackage{times}    
\usepackage{amsmath}  
\usepackage{amssymb}  
\usepackage{savesym} \savesymbol{AND}
\usepackage{algorithm,algorithmic}
\usepackage{multirow}
\usepackage{natbib}   
\usepackage{cancel}

\input{mathdef}

\begin{document}
\begin{frontmatter}

\title{Sequential Linear Quadratic Optimal Control for Nonlinear Switched Systems} 

\thanks[footnoteinfo]{This research has been supported in part by a Max-Planck ETH Center for Learning Systems Ph.D. fellowship to Farbod Farshidian, a Swiss National Science Foundation Professorship Award to Jonas Buchli, the NCCR Robotics, and Maryam Kamgarpour's ERC Starting Grant CONENE.}

\author[adrl]{Farbod Farshidian} 
\author[ifa]{Maryam Kamgarpour}
\author[adrl]{Diego Pardo}
\author[adrl]{Jonas Buchli} 

\address[adrl]{Agile and Dexterous Robotics Lab, ETH Z\"urich, Switzerland \\(e-mail: \{farbodf, depardo, buchlij\}@ethz.ch),}
\address[ifa]{Automatic Control Laboratory, ETH Z\"urich, Switzerland \\(e-mail: mkamgar@control.ee. ethz.ch).}

\begin{abstract}                
In this contribution, we introduce an efficient method for solving the optimal control problem for an unconstrained nonlinear switched system with an arbitrary cost function. We assume that the sequence of the switching modes are given but the switching time in between consecutive modes remains to be optimized. The proposed method uses a two-stage approach as introduced by \cite{xu04} where the original optimal control problem is transcribed into an equivalent problem parametrized by the switching times and the optimal control policy is obtained based on the solution of a two-point boundary value differential equation. The main contribution of this paper is to use a Sequential Linear Quadratic approach to synthesize the optimal controller instead of solving a boundary value problem. The proposed method is numerically more efficient and scales very well to the high dimensional problems. In order to evaluate its performance, we use two numerical examples as benchmarks to compare against the baseline algorithm. In the third numerical example, we apply the proposed algorithm to the Center of Mass control problem in a quadruped robot locomotion task.
\end{abstract}


\end{frontmatter}

\section{Introduction}
Switched systems are a subclass of a general family known as hybrid systems. Hybrid system model consists of a finite number of dynamical subsystems subjected to discrete events which cause transition between these subsystems. This transition is either triggered by an external input, or through the intersection of the continuous states trajectory to a certain manifolds known as the switching surfaces. Switched systems are usually characterized by systems that have continuous state transition during these switches. Switched system models are encountered in many practical applications such as automobiles and locomotives with different gears, DC-DC converters, manufacturing processes, biological systems, and robotics. 

Our interest in switched systems originates from an application on a legged robot where we model the Center of Mass (CoM) as a switched system. The control goal is to synthesize a controller which, for a given gait, stabilizes the robot while it minimizes a cost function. To fulfill this task, the robot can manipulate the ground reaction forces at the stance feet and adjust the switching times between different stance leg configurations. For instance, assume the problem of controlling the walking gait for a quadruped robot. In this task, the gait is fixed, thus the sequence of mode switches are known. The control task is to modulate the contact forces of the stance legs and to determine the switching times between each mode. 

The optimal control problem for the switched systems involves synthesizing the optimal controller for the continuous inputs and finding a mode sequence and the switching times between the modes. In general, the procedure of synthesizing an optimal control law for a switched system can be divided into three subtasks \citep{giua01, xu04}: (1) finding the optimal sequence of the modes, (2) finding the optimal switching times between consecutive modes, (3) finding the optimal continuous control inputs. Given the switching sequence and times, the third subtask is a regular optimal control problem with a finite number of discontinuities in the system vector field. The necessary condition of optimality in the context of hybrid systems has been derived from Pontryagin's maximum principle \citep{branicky98, sussmann99, riedinger03} and subsequently, various computational techniques have been developed to solve this problem \citep{shaikh07, soler12, pakniyat14}. 

Based on the Pontryagin's maximum principle, the optimal solution should satisfy a two-point boundary value problem (BVP). However, similar to the classical control problem, the difficulties related to numerical solution of the necessary condition of optimality limits the application of this approach. In \cite{riedinger99}, it has been shown that for a Linear-Quadratic (LQ) problem it is sufficient to solve a sequence of Riccati equations with proper transversality conditions at the switching times in order to optimize the continuous inputs but the mode switches should be  still calculated based on the enumerations of all the possible switches at each time step. In order to ease the computational burden of finding the optimal switching behavior, in \cite{bengea05}, the switched system is embedded in larger family of systems defined by the convex hull of the switched subsystems. It has been shown that if the sufficient and necessary conditions for optimality in the embedded systems exists, the bang-bang optimal solution of the embedded problem is also the optimal solution of the switched system; otherwise a sub-optimal solution can be derived. 

\cite{borrelli05} propose an off-line method to synthesize an optimal control law for a discrete linear hybrid system with linear inequality constraints. The proposed method is a combination of dynamic programming and multi-parametric quadratic programming which designs a feedback law for continuous and discrete inputs in the feasible regions. A simpler approach in \cite{bemporad99} uses a mixed integer linear/ quadratic program to solve the optimal control problem for mixed logical dynamical systems. 

Optimizing the cost function with respect to the switching times has been studied for autonomous systems by \cite{egerstedt03, johnson11, wardi12} and for non-autonomous systems by \cite{kamgarpour12}. By using the derivative of the cost function with respect to the switching time, these methods use nonlinear programming techniques to optimize the cost function. However, in general these methods do not consider the sensitivity of the continuous inputs' control law to the switching times.  

While many of the aforementioned approaches are computationally demanding for real-time robotic applications, there is a class of efficient optimal control algorithms known as Sequential Linear Quadratic (SLQ) methods which can be applied to real-time, complex robotic applications \citep{neunert16}. An SLQ algorithm sequentially solves the extremal problem around the latest estimation of the optimal trajectories and improves these optimal trajectories using the extremal problem solution \citep{mayne66,todorov05,sideris05}. Motivated by their efficiency in solving regular optimal control problems, in this paper we have extended an SLQ algorithm to solve the optimal control problem for nonlinear switched system with predefined mode sequence. To this end, we adopt an approach introduced by \cite{xu04} where the primary switched problem is transcribed into an equivalent problem. We then introduce a two-stage optimization method to optimize the continuous inputs and the switching times. While \cite{xu04} use a computationally demanding approach based on solving a set of two-point BVPs, we propose a new SLQ algorithm to efficiently solve the optimal control problem for nonlinear switched systems.  

The main contributions of this paper are: (1) it uses an efficient SLQ algorithm to synthesize the optimal control law for the continuous inputs. (2) it calculates the cost function derivative with respect to switching times using an LQ approximation of the problem. This approximation is obtained without any additional computation from the SLQ solution. (3) it introduces a new practical application of the optimal control for switched systems in the field of motion planning of legged robots.

\section{Problem Formulation} \label{sec:problem} \vspace{-1mm}
In this section, we briefly introduce the optimal control problem based on the parameterization of switching times. We assume that the switched system dynamics consist of $I$ subsystems where the system dynamics for the $i$th subsystem (${ i \in \{1,2,\dots,I\} }$) is as follows
\begin{equation} \label{eq:system_dynamics}
\dot{\vx}(t) = \vf_i \left( \vx(t),\vu(t) \right) \qquad \text{\textit{for }} \, t_{i-1} \leq t < t_{i},
\end{equation}
where $\vx(t) \in \mathbb{R}^{n_x}$ is the continuous state, $\vu(t) \in \mathbb{R}^{n_u}$ is the piecewise continuous control input, and $\vf_i: \mathbb{R}^{n_x} \times \mathbb{R}^{n_u} \rightarrow \mathbb{R}$ is the vector field of subsystem $i$ which is continuously differentiable and Lipschitz  up to the first order derivatives. $t_i$ is the switching time between subsystem $i$ and $i+1$. $t_0$ and $t_I$ are respectively the given initial time and the final time. The initial state is $\vx_0$ and $\vx(t_{i}^-) = \vx(t_{i}^+)$ at the switching moments because of the state continuity condition. The optimal control problem for the switched system in Equation~\eqref{eq:system_dynamics} is defined as
\begin{equation} \label{eq:general_opt}
\min\limits_{\scriptstyle \begin{matrix}
\vt\!\in\!\mathbb{T}, \vu(\cdot)
\end{matrix}
} \Phi(\vx(t_I))+ \sum_{i=1}^I{\int_{t_{i-1}}^{t_{i}} { L_i(\vx,\vu)dt}},
\end{equation}
where $\Phi(\cdot)$ and $L_i(\cdot,\cdot)$ are the final cost and the running cost (for subsystem $i$) which are continuously differentiable and Lipschitz up to the second order derivatives. $\mathbb{T}$ is a polytope in $\mathbb{R}^{^{I-1}}$ defined as
$\mathbb{T} = \left\{ (t_1, \dots, t_{I-1}) | t_o \leq t_1 \leq \dots \leq t_{I-1} \leq t_I \right\}$. 

The optimal control problem based on the parameterization of switching times can be defined as the following two-stage optimization problem 
\begin{align} \label{eq:two_stage_opt}
\min_{\vt \in \mathbb{T}} J[\vt,\vx^*,\vu^*] \quad \text{s.t. } \{\vx^*,\vu^*\} = \argmin J[\vt,\vx,\vu]. 
\end{align}
with
\begin{align}
J[\vt,\vx,\vu] &= \Phi(\vx_{I})+ \sum_{i=1}^I{(t_{i}-t_{i-1}) \int_{i-1}^{i} {L_i(\vx(z),\vu(z))dz}}, \label{eq:equivalent_cost} \\
\frac{d\vx(z)}{dz} &= (t_{i}-t_{i-1}) \vf_i \left(\vx(z),\vu(z) \right) \qquad \text{\textit{for }} i-1 \leq z < i 
\label{eq:equivalent_system} \\
t &= (t_{i}-t_{i-1})(z-i) + t_{i}, \label{eq:z}
\end{align}   
in which we have replaced the independent time variable $t$ with a normalized time variable $z$ defined by \eqref{eq:z}. With this change of variable, while the switching times are still part of the decision variables, they are fixed parameters for the bottom-level optimization in \eqref{eq:two_stage_opt}. Therefore, the reformulated optimal control problem does not have variable switching times and it reduces to a conventional optimal control problem parameterized over the switching times (Theorem 1 in \cite{xu04}). In the next section, we introduce our algorithm for calculating the optimized cost function and its gradient.

\section{Solution Approach} \label{sec:ocs2}
\vspace{-3mm}
In this section, we use a gradient-based method to solve the top-level optimization introduced in \eqref{eq:two_stage_opt} and a dynamic programming approach to synthesize the continuous inputs control law. In each iteration of the gradient-based method for finding the optimal switching time, we first solve a continuous-time optimal control problem for the system with a fixed switching times. Then, the cost function gradient with respect to the switching times is calculated in order to determine the descent direction for the switching times update. This approach is similar to \cite{xu04}. However, instead of using the two-point BVP solver for optimizing $J(\vt)$ with respect to $\vu$ and calculating its gradient with respect to $\vt$, we use a more efficient approach. This facilitates implementing this algorithm in real-time on complex systems such as legged robots. 

Our \textbf{O}ptimal \textbf{C}ontrol for \textbf{S}witched \textbf{S}ystems algorithm (OCS2 algorithm) consists of two main steps: a method which synthesizes the continuous input controller and a method which calculates the parameterized cost function derivatives with respect to the switching times. For synthesizing the continuous input controller, OCS2 uses the SLQ algorithm in Algorithm~\ref{alg:slq}. As a dynamic programming approach, SLQ uses the Bellman equation of optimality to locally estimate the value function and consequently the optimal control law. In the second step of OCS2, we use the approximated problem for calculating the value function from the first step to efficiently compute the value function gradient with respect to switching times.

Using the SLQ algorithm to calculate the optimal control law for the continuous inputs has two major advantages. First, as discussed by \cite{sideris10}, this algorithm has a linear time complexity with respect to the optimization time horizon in contrast to many standard discretization-based algorithms which scale cubically (such as the direct collocation methods). Second, since in each iteration of SLQ, an LQ problem is optimized, we can efficiently calculate the value function derivatives by obtaining derivatives of an LQ subproblem's value function. However, in contrast to a regular time-variant LQ problem in which the system dynamics and the cost function are independent of the switching times, in this problem the cost and the system dynamics of the approximated LQ subproblem are functions of the switching times. Therefore, in order to calculate the cost function gradient, we should also consider the LQ subproblem variations to the switching times. We tackle this problem more rigorously in Theorem 1.  

Here, we first briefly introduce the SLQ algorithm. We consider an intermediate iteration of the algorithm. We assume that $\{\bar{\vx}(z)\}_{z= 0}^{I}$ and $\{\bar{\vu}(z)\}_{z= 0}^{I}$ are the nominal state and input trajectories which are obtained by forward integrating system dynamics in \eqref{eq:equivalent_system} (performing a rollout) using the latest estimation of the optimal control law with a fixed switching time vector $\bar{\vt}$. For simplicity of notation, in the followings we have dropped the dependencies of the nominal trajectories and consequently the approximated LQ problem with respect to $\bar{\vt}$. The linearized system dynamics of the equivalent system in \eqref{eq:equivalent_system} around these nominal trajectories are defined as      
\begin{align} 
& \cfrac{d(\delta\vx)}{dz} = (t_{i}-t_{i-1}) \left( \vA_i(z)\delta\vx + \vB_i(z)\delta\vu \right) \quad  i-1 \leq z < i  \notag \\
& \vA_i(z) = \frac{\partial \vf_i(\bar{\vx}(z),\bar{\vu}(z))}{\partial\vx}, \quad \vB_i(z) = \frac{\partial \vf_i(\bar{\vx}(z),\bar{\vu}(z))}{\partial\vu} ,
\label{eq:dynamics_linear_approximation}
\end{align}
where $\delta \vx(z) = \vx(z)-\bar{\vx}(z)$, $\delta\vu(z) = \vu(z)-\bar{\vu}(z)$. The quadratic approximation of the cost function in \eqref{eq:equivalent_cost} is
\begin{align}
& \tilde{J} = \tilde{\Phi}(\vx_{t_f})+ \sum_{i=1}^I{\int_{i-1}^{i} { (t_{i}-t_{i-1})\tilde{L}_i(z,\vx,\vu)dz}} \notag \\
&\tilde{\Phi}_f(\vx) = \ q_f + \vq_f^\top \delta\vx + \frac{1}{2} \delta\vx^\top \vQ_f \delta\vx \notag \\
& \tilde{L}_i(z,\vx,\vu) = q_i(z) + \delta\vx^\top \vq_i(z) + \delta\vu^\top \vr_i(z)  + \delta\vx^T \vP_i(z) \delta\vu \notag \\
& \hspace{15mm} + \frac{1}{2} \delta\vx^\top \vQ_i(z) \delta\vx + \frac{1}{2} \delta\vu^\top \vR_i(z) \delta\vu.  \label{eq:cost_quadratic_approximation}
\end{align} 
In the above, $q(z)$, $\vq(z)$, $\vr(z)$, $\vQ(z)$, $\vP(z)$, and $\vR(z)$ are the coefficients of the Taylor expansion of the cost function in \eqref{eq:equivalent_cost} evaluated at the nominal trajectories. The optimal control law for this LQ extremal subproblem can be derived by solving the following Riccati equations \citep{bryson75}
\begin{align}
& -\frac{d\vS(z)}{dz}  = (t_{i}-t_{i-1}) \vW(z), \hspace{3mm}   \vS(i^-)=\vS(i^+), \hspace{2mm} \vS(I)=\vQ_{f} \label{eq:riccati_Sm}  \\
& -\frac{d\vs(z)}{dz}  = (t_{i}-t_{i-1}) \vw(z), \hspace{5mm}   \vs(i^-)=\vs(i^+), \hspace{2mm} \vs(I)=\vq_{f} \label{eq:riccati_Sv} \\
& -\frac{ds(z)}{dz}    = (t_{i}-t_{i-1}) w(z),  \hspace{5mm}   s(i^-)=s(i^+), \hspace{2mm} s(I)=q_{f}   \label{eq:riccati_s} ,
\end{align}
where $\vS(z)$ and $\vW(z)$ are in $\mathbb{R}^{^{n_x \times n_x}}$, $\vs(z)$ and $\vw(z)$ are in $\mathbb{R}^{^{n_x}}$, $s(z)$ and $w(z)$ are in $\mathbb{R}$. These matrices are defined as
\begin{align}
&\vW(z) = \vQ(z) + \vA(z)\!^\top \vS(z) + \vS(z) \vA(z) - \vL(z)\!^\top \vR(z) \; \vL(z) \label{eq:riccati_Sm_o}  \\
& \vw(z)  = \vq(z) + \vA(z)^\top \vs(z) - \vL(z)^\top \vR(z) \; \vl(z) \label{eq:riccati_Sv_o}  \\
& w(z) = q(z) - 0.5\alpha (2-\alpha)\; \vl(z)^\top \vR(z) \; \vl(z)  \label{eq:riccati_s_o} \\
&\vl(z) = - \vR(z)^{-1} \big( \vr(z) + \vB(z)^\top \vs(z) \big) \label{eq:optimal_control_l}\\
&\vL(z) = - \vR(z)^{-1} \big( \vP(z)^\top + \vB(z)^\top \vS(z) \big) . \label{eq:optimal_control_L} 
\end{align}
The updated optimal control law, the value function, and the total cost are defined as
\begin{align}
& \vu(z,\vx) = \bar{\vu}(z) + \alpha \vl(z) + \vL(z) \delta\vx \label{eq:slq_policy}\\
& V(z,\vx) = s(z) + \delta\!\vx^\top \vs(z) + \frac{1}{2}\delta\!\vx^\top \vS(z) \delta\!\vx  \label{eq:value_function} \\
& \tilde{J} = V(0,\vx_0) = s(0),
\end{align}
where $z$ is defined in \eqref{eq:z}. $\alpha \in [0,1]$ is the learning rate for backtracking line-search \citep{armijo96}. In each iteration of SLQ, the line-search parameter controls the maximum step to move along the feedforward component of the control law update. This parameter is chosen by starting with a full step in the direction of the update ($\alpha=1$), and then iteratively shrinking the step size until the cost associated to the updated controller rollout is lower than the current nominal trajectories cost. Before proceeding to the main theorem and its proof, we state the following lemma which will be used in Theorem~1.   
\begin{lem}
$H: \mathbb{R}^n \rightarrow \mathbb{R}$ and $\vy: \mathbb{R} \rightarrow \mathbb{R}^n$ as twice continuously differentiable functions. If $|\tau-\bar{\tau}|<\varepsilon_1$ and $|\delta\!\vy|<\varepsilon_2$, for small enough $\varepsilon_1$ and $\varepsilon_2$, $H(\vy(\tau)+\delta\!\vy)$ can be approximated as
\begin{align}
H(\vy(\tau)+\delta\!\vy) &\simeq H(\bar{\vy}) + \delta\!\vy^\top \nabla\!H(\bar{\vy}) + \partial\bar{\vy}^\top \nabla\!H(\bar{\vy}) (\tau-\bar{\tau}) \notag \\
& + \frac{1}{2} \delta\!\vy^\top \nabla^2\!H(\bar{\vy}) \delta\!\vy + \delta\!\vy^\top \nabla^2\!H(\bar{\vy}) \partial\bar{\vy} (\tau-\bar{\tau})  \notag \\
& + \frac{1}{2} \Big( \partial\bar{\vy}^\top \nabla^2\!H(\bar{\vy}) \partial\bar{\vy} + \partial^2\bar{\vy}^\top \nabla\!H(\bar{\vy}) \Big) (\tau-\bar{\tau})^2 , \notag
\end{align}
where $\bar{\vy} = \vy(\bar{\tau})$, $\partial\bar{\vy} = \frac{d\vy}{d\tau}|_{\bar{\tau}}$, and $\partial^2\bar{\vy} = \frac{d^2\vy}{d\tau^2}|_{\bar{\tau}}$. Furthermore, $\nabla\!H$ and $\nabla^2\!H$ are the gradient and the Hessian of $H$ respectively. 
\end{lem}

\begin{pf}
This lemma can be easily proven using the Taylor series expansion of $H$ and few steps of simplification.  
\end{pf}

\begin{thm} The partial derivative of the value function in \eqref{eq:value_function} and the partial derivative of the total cost with respect to the switching time $t_j$ can be derived as 
\begin{align}
\partial_{t_j}V(z,\vx) =& \partial_{t_j}s(z) + \delta\!\vx^\top \partial_{t_j}\vs(z) + \frac{1}{2}\delta\!\vx^\top \partial_{t_j}\vS(z) \delta\!\vx  - \partial_{t_j}\bar{\vx}^\top \vs(z) \notag \\
& - \frac{1}{2} \partial_{t_j}\bar{\vx}^\top \vS(z) \delta\!\vx - \frac{1}{2} \delta\!\vx^\top \vS(z) \partial_{t_j}\bar{\vx}   \\
\partial_{t_j}\tilde{J} =& \partial_{t_j}s(0),
\end{align}
where $z$ is defined by \eqref{eq:z}, $\partial_{t_j}$ is the partial derivative operator with respect to switching time $t_j$. Then $\partial_{t_j}\bar{\vx}$ and $\partial_{t_j}\bar{\vu}$ are respectively the state and input sensitivity to switching time $t_j$ which are calculated by the following differential equation
\begin{align}
\frac{d (\partial_{t_j}\bar{\vx})}{dz} =& ( \delta_{i,j} - \delta_{i-1,j} ) \vf_i(\bar{\vx}(z),\bar{\vu}(z)) \notag \\
& + (t_{i}-t_{i-1}) \left( \vA_i(z) \partial_{t_j}\bar{\vx} + \vB_i(z) \partial_{t_j}\bar{\vu} \right)  \label{eq:state_sensitivity} \\
\partial_{t_j}\bar{\vu}(z,\partial_{t_j}\bar{\vx}) &= -\vL(z) \partial_{t_j}\bar{\vx} + \partial_{t_j}\vL(z) \delta\bar{\vx} + \partial_{t_j}\vl(z) ,
\label{eq:input_sensitivity}
\end{align}
with the initial condition $\partial_{t_j}\bar{\vx}(0) = \mathbf{0}$ and the transversality condition $\partial_{t_j}\bar{\vx}(i^-) = \partial_{t_j}\bar{\vx}(i^+)$. $\delta_{i,j}$ is the Kronecker delta which is one if the variables are equal, and zero otherwise. Furthermore, $\partial_{t_j}\vS(z)$, $\partial_{t_j}\vs(z)$, and $\partial_{t_j}s(z)$ can be calculated from the following set of linear differential equations
\begin{align}
-\frac{d (\partial_{t_j}\vS(z))}{dz}  &= ( \delta_{i,j} - \delta_{i-1,j} ) \vW(z) + (t_{i}-t_{i-1}) \partial_{t_j}\!\vW(z)  \label{eq:riccati_nablaSm}  \\
-\frac{d (\partial_{t_j}\vs(z))}{dz}  &= ( \delta_{i,j} - \delta_{i-1,j} ) \vw(z) + (t_{i}-t_{i-1}) \partial_{t_j}\!\vw(z)  \label{eq:riccati_nablaSv} \\
-\frac{d (\partial_{t_j}s(z))}{dz}    &= ( \delta_{i,j} - \delta_{i-1,j} ) w(z) + (t_{i}-t_{i-1}) \partial_{t_j}\!w(z) , \label{eq:riccati_nablas}
\end{align}
where we have
\begin{align*}
&\partial_{t_j}\!\vW(z) = \vA(z)^\top \partial_{t_j}\vS(z) + \partial_{t_j}\vS(z)  \vA(z) - \partial_{t_j}\vL(z)^\top \vR(z) \vL(z) \notag \\
& \hspace{14mm} - \vL(z)^\top \vR(z) \partial_{t_j}\vL(z)  \\
&\partial_{t_j}\!\vw(z) = \partial_{t_j}\vq(z) + \vA(z)^\top \partial_{t_j}\vs(z) - \partial_{t_j}\vL(z)^\top \vR(z) \vl(z) \notag \\
& \hspace{14mm} - \vL(z)^\top \vR(z) \partial_{t_j}\vl(z) \\
&\partial_{t_j}\!w(z)  = \partial_{t_j}q(z) - 0.5\alpha (2-\alpha) \big( \partial_{t_j}\vl(z)^\top \vR(z) \vl(z)  \notag \\
& \hspace{14mm} + \vl(z)^\top \vR(z) \partial_{t_j}\vl(z) \big) \\
&\partial_{t_j}\vl(z) = - \vR(z)^{-1} \big( \partial_{t_j}\vr(z) + \vB(z)^\top \partial_{t_j}\vs(z) \big)  \\
&\partial_{t_j}\vL(z) = - \vR(z)^{-1} \vB(z)^\top \partial_{t_j}\vS(z) .
\end{align*}
The derivative of the cost coefficients with respect to $t_j$ is
\begin{equation}
\begin{bmatrix} \partial_{t_j}\vq(z) \\ \partial_{t_j}\vr(z) \\ \partial_{t_j}q(z) \end{bmatrix} = 
\begin{bmatrix} \vQ(z) & \vP(z) \\ \vP(z)^\top & \vR(z) \\ \vq(z)^\top & \vr(z)^\top \end{bmatrix}
\begin{bmatrix} \partial_{t_j}\vx(z) \\ \partial_{t_j}\vu(z) \end{bmatrix} .  
\label{eq:intermidiate_cost_function_sensitivity}
\end{equation}
Finally, the terminal condition are defined as
\begin{equation}
\partial_{t_j}\!\vS(I)=\mathbf{0}, \hspace{2mm} \partial_{t_j}\!\vs(I)=\vQ_f \partial_{t_j}\!\vx(I), \hspace{2mm} \partial_{t_j}\!s(I)=\vq_f^\top \partial_{t_j}\!\vx(I),
\label{eq:final_cost_function_sensitivity}
\end{equation}
and the transversality conditions are 
\begin{equation*}
\partial_{t_j}\!\vS(i^-) = \partial_{t_j}\!\vS(i^+),  \hspace{2mm} 
\partial_{t_j}\!\vs(i^-) = \partial_{t_j}\!\vs(i^+),  \hspace{2mm}
\partial_{t_j}\!s(i^-) = \partial_{t_j\!}s(i^+).
\end{equation*}
\end{thm}

\begin{algorithm}[t]
\caption{SLQ Algorithm}
\label{alg:slq}
\begin{algorithmic}
	\STATE \textbf{Given}
	\STATE - Mode switch sequence and switching times
	\STATE - Cost function \eqref{eq:equivalent_cost} and system dynamics \eqref{eq:equivalent_system} 
	\STATE \textbf{Initialization}
	\STATE - Initialize the controller with a stable control law, $\vu_0(\cdot)$
	\REPEAT
	\STATE - Forward integrate system dynamics: $\{\ \bar{\vx}(z),\bar{\vu}(z)\}_{z=0}^{I},$
	\STATE - Compute the LQ approximation of the problem along the nominal trajectory, \eqref{eq:dynamics_linear_approximation} and \eqref{eq:cost_quadratic_approximation}.
	\STATE - Solve the final value differential equations, (\ref{eq:riccati_Sm}-\ref{eq:riccati_s}).
	\STATE - Line search for the optimal $\alpha$ with policy \eqref{eq:slq_policy}
	\STATE - Update control law: $\vu^*(z,\vx) = \bar{\vu}(z) + \alpha^* \vl(z) + \vL(z) \delta\vx $
	\UNTIL{$\left\Vert \vl(\cdot) \right\Vert_2 < l_{min}$ or maximum number of iterations}
\end{algorithmic}
\end{algorithm} 

\begin{algorithm}[t]
\caption{OCS2 Algorithm}
\label{alg:ocs2}
\begin{algorithmic}
	\STATE \textbf{Given}
	\STATE - Mode switch sequence and initial switching times, $\vt_0$
	\STATE - The optimal control problem in equations \eqref{eq:system_dynamics} and  \eqref{eq:general_opt}  
	\STATE \textbf{Initialization}
	\STATE - Empty the solution bag
	\STATE - Initialize SLQ policy, $\vu_0(\cdot)$
	\STATE - Initialize the switching times, $\vt_0$
	\REPEAT
	\STATE - Compute the equivalent cost function in Equation \eqref{eq:equivalent_cost}
	\STATE - Compute the equivalent system dynamics in Equation \eqref{eq:equivalent_system} 
	\IF{Solution bag in not empty} 
		\STATE - Initialized the SLQ policy with the controller in the solution bag that has the most similar switching vector
	\ELSE 
		\STATE - Initialized the SLQ policy, $\vu_0(z,\vx)$.
	\ENDIF 
	\STATE - Run the SLQ algorithm
	\STATE - Get the optimal control $\vu^*(z,\vx;\vt^*)$ 
	\STATE - Memorize the pair $\left(\vt^*,\vu^*(z,\vx;\vt^*)\right)$ in the solution bag
	\STATE - Calculate $\partial_{t_j}\vS(0)$, $\partial_{t_j}\vs(0)$, and $\partial_{t_j}s(0)$ using  (\ref{eq:riccati_nablaSm}-\ref{eq:riccati_nablas}).
	\STATE - Calculate the cost function gradient $\nabla_{\vt}J = [\partial_{t_j}s(0)]_j$
	\STATE - Use a gradient-descent method to update $\vt^*$ 
	\UNTIL{gradient-descent method converges} \\
	\STATE \textbf{Return} the optimal $\vt^*$, and $\vu^*(z,\vx;\vt^*)$.
\end{algorithmic}
\end{algorithm} 

\begin{pf}
Equations (\ref{eq:riccati_nablaSm}-\ref{eq:riccati_nablas}) can be derived by directly differentiating the Riccati equations in (\ref{eq:riccati_Sm}-\ref{eq:riccati_s}), where based on the continuity condition, the order of differentiation with respect to time, $z$, and $t_j$ has been changed. For calculating $\partial_{t_j}\!\vW$, $\partial_{t_j}\!\vw$, and $\partial_{t_j}\!w$, we need to calculate the derivative of the linearized dynamics and the quadratic approximation of the cost function with respect to switching time $t_j$ as well. To do so, we need to examine the impact of $t_j$ variation on the system dynamics and cost function of the  approximated LQ subproblem.

In each approximated LQ subproblem, we use a quadratic approximation of the cost function components $L$ and $\Phi$ around the nominal trajectories $\bar{\vx}(z)$ and  $\bar{\vu}(z)$ of the equivalent system \eqref{eq:equivalent_system}. We can readily derive the differential equation which determines the sensitivity of the state trajectory with respect to switching time $t_j$ by differentiating both side of the equivalent systems dynamics in \eqref{eq:equivalent_system} with respect to $t_j$.
\begin{equation*}
\partial_{t_j} \frac{d \bar{\vx}}{dz} = ( \delta_{i,j} - \delta_{i-1,j} ) \vf_i(\bar{\vx},\bar{\vu}) \notag + (t_{i}-t_{i-1}) \left( \vA_i \partial_{t_j}\!\bar{\vx} + \vB_i \partial_{t_j}\!\bar{\vu} \right).
\end{equation*}
For simplicity in notation, we drop this dependencies on time, $z$. Using the continuity condition of the state trajectory, the order of derivatives on the right hand side of the equation is changed which is resulted in \eqref{eq:state_sensitivity}. Furthermore, since the control input trajectory in SLQ comprises a time-varying feedforward term and a time-varying linear state feedback, its sensitivity can be calculated as \eqref{eq:input_sensitivity}. Moreover, for the initial condition of this equation we have $\partial_{t_j}\bar{\vx}(z=0) = \mathbf{0}$
due to the fixed initial state. Based on Lemma 1, the second order approximation of the intermediate cost $L(\vx,\vu)$ around the nominal trajectories (which are in turn a function of $t_j$) can be written as
\begin{align}
& L(z,\bar{\vx}(s)+\delta\!\vx,\bar{\vu}(s)+\delta\!\vu) \simeq 
q(z) +
\begin{bmatrix} \vq(z) \\ \vr(z) \end{bmatrix}^\top 
\begin{bmatrix} \partial_{t_j}\vx(z) \\ \partial_{t_j}\vu(z) \end{bmatrix} (t_j-\bar{t}_j)
\notag  \\
& \hspace{2mm} 
+\begin{bmatrix} \delta\!\vx \\ \delta\!\vu \end{bmatrix}^\top\! 
\left(
\begin{bmatrix} \vq(z) \\ \vr(z) \end{bmatrix} +
\begin{bmatrix} \vQ(z) & \vP(z) \\ \vP(z)^\top & \vR(z) \end{bmatrix}
\begin{bmatrix} \partial_{t_j}\vx(z) \\ \partial_{t_j}\vu(z) \end{bmatrix} 
(t_j-\bar{t}_j) \right) \notag \\
& \hspace{2mm} 
+\frac{1}{2}
\begin{bmatrix} \delta\!\vx \\ \delta\!\vu \end{bmatrix}^\top\!\! 
\begin{bmatrix} \vQ(z) & \vP(z) \\ \vP(z)^\top & \vR(z) \end{bmatrix}
\begin{bmatrix} \delta\!\vx \\ \delta\!\vu \end{bmatrix}
+ \frac{(t_j-\bar{t}_j)^2}{2}
\begin{bmatrix} \vq(z) \\ \vr(z) \end{bmatrix}^\top\! 
\begin{bmatrix} \partial_{t_j}^2\vx(z) \\ \partial_{t_j}^2\vu(z) \end{bmatrix}
\notag \\  
& \hspace{2mm} 
+ \frac{(t_j-\bar{t}_j)^2}{2}
\begin{bmatrix} \partial_{t_j}\vx(z) \\ \partial_{t_j}\vu(z) \end{bmatrix}^\top\! 
\begin{bmatrix} \vQ(z) & \vP(z) \\ \vP(z)^\top & \vR(z) \end{bmatrix}
\begin{bmatrix} \partial_{t_j}\vx(z) \\ \partial_{t_j}\vu(z) \end{bmatrix}
\end{align}
    
In the above, we have combined the like terms of the state and input increments. Based on this equation the sensitivity of the cost function components will be
\begin{align*}
\partial_{t_j} \vr(z) =&  \vP(z)^\top \partial_{t_j}\vx(z) + \vR(z)  \partial_{t_j}\vu(z), \hspace{3mm} \partial_{t_j} \vR(z) = \mathbf{0}, \\
\partial_{t_j} \vq(z) =&  \vQ(z) \partial_{t_j}\vx(z) + \vP(z) \partial_{t_j}\vu(z), \hspace{5mm} \partial_{t_j} \vQ(z) = \mathbf{0}, \\
\partial_{t_j} q(z)   =& \vq(z)^\top \partial_{t_j}\vx(z) + \vr(z)^\top \partial_{t_j}\vu(z), \hspace{2mm} \partial_{t_j} \vP(z) = \mathbf{0},
\end{align*}
which can be written in matrix form as \eqref{eq:intermidiate_cost_function_sensitivity}. By the same process, we can derive \eqref{eq:final_cost_function_sensitivity} for the final cost. 

In order to find the linearized system dynamics sensitivity with respect to the switching time $t_j$, we can use the result from Lemma 1 where we only keep the terms up to the first order.
\begin{align*}
& \frac{d}{dz} \left( \bar{\vx} + \partial_{t_j}\!\vx(t_j-\bar{t}_j) + \delta\!\vx \right) \simeq (\bar{t}_i-\bar{t}_{i-1}) \big( \vf(\bar{\vx},\bar{\vu}) + \vA_i \delta\!\vx + \vB_i \delta\!\vu \big) \notag \\
&  \hspace{0mm}
 + (t_j-\bar{t}_j) \big( ( \delta_{i,j} - \delta_{i-1,j} ) \vf_i(\bar{\vx},\bar{\vu}) + (\bar{t}_{i}-\bar{t}_{i-1}) \left( \vA_i \partial_{t_j}\bar{\vx} +  \vB_i \partial_{t_j}\bar{\vu} \right) \big) ,
\end{align*}
where for simplicity in notation, we drop the dependencies on time, $z$. By equating the coefficients, we get
\begin{align*}
&\frac{d \bar{\vx}}{dz} \simeq (\bar{t}_i-\bar{t}_{i-1})f(\bar{\vx},\bar{\vu}) \\
&\frac{d \delta\!\vx}{dz} \simeq 
(\bar{t}_{i}-\bar{t}_{i-1}) \big( \vA_i \delta\!\vx + \vB_i \delta\!\vu \big) \\
&\frac{d (\partial_{t_j}\vx)}{dz} \simeq ( \delta_{i,j} - \delta_{i-1,j} ) \vf_i(\bar{\vx},\bar{\vu}) + (\bar{t}_{i}-\bar{t}_{i-1}) \big(\vA_i \partial_{t_j}\!\bar{\vx} + \vB_i \partial_{t_j}\!\bar{\vu} \big) ,
\end{align*}
which are respectively the nominal trajectory equation, the linear approximation of system dynamics, and trajectory sensitivity. Based on this approximation, the linear part is not a function of $t_j$, so we get $\partial_{t_j}\!\vA_i(z) = \mathbf{0}$ and $\partial_{t_j}\!\vB_i(z) = \mathbf{0}$ (note that the effect of the switching time variations on the approximated system dynamics would have been appeared if we had used a second order or a higher order approximation). 
\end{pf} 

\vspace{-2mm}
\section{OCS2 algorithm}
\vspace{-3mm}
In Section~\ref{sec:ocs2}, we have discussed the technical details behind the OCS2 algorithm. Here, we highlight the main steps of the algorithm for synthesizing the optimal continuous control law and the optimal switching times (refer to Algorithm \ref{alg:ocs2}). Each iteration of the OCS2 algorithm has three main steps namely (1) using SLQ algorithm to find the continuous inputs' optimal control, (2) calculating the cost function gradient based on the LQ approximation of the problem which has been already calculated by the SLQ algorithm, (3) using a gradient descent method to update the switching times where we use the Frank-Wolfe method \citep{jaggi13}. An interesting aspect of our algorithm is its linear-time computational complexity with respect to the optimization time-horizon both in SLQ algorithm and calculating cost gradient. As we will show in the next section, this characteristic results in a superior performance of our algorithm in comparison to the two-point BVP approach originally introduced by \cite{xu04}.   

As discussed, OCS2 uses the SLQ algorithm for solving a nonlinear optimal control problem on the equivalent system with fixed switching times (refer to Algorithm \ref{alg:slq}). The SLQ algorithm is based on an iterative scheme where in each iteration it forward integrates the controlled system and then approximates the nonlinear optimal control problem with a local LQ problem. In general, the SLQ algorithm requires an initial stable controller for the first forward integration. For deriving the initial controller, we define a set of operating points in each switching mode (normally one point) and approximate the optimal control problem around these operating points with an LQ approximation. Then, we follow the same process described in the SLQ algorithm to design an initial controller.  

A good initialization can often increase the convergence speed of the SLQ algorithm. One interesting characteristic of our algorithm is the warm starting scheme for the initial policy of SLQ. Here, we use a memorization scheme where we store the solutions of the different runs of SLQ in a solution bag and later initialize the policy of a new run of SLQ with the most similar switching time's policy (refer to Algorithm \ref{alg:ocs2}). The similarity between two switching time sequences is measured as a sum of squared differences between corresponding times in two sequences.

\section{Case Studies} \label{sec:numerical_example}
\begin{figure}[t]
    \includegraphics[width=\columnwidth]{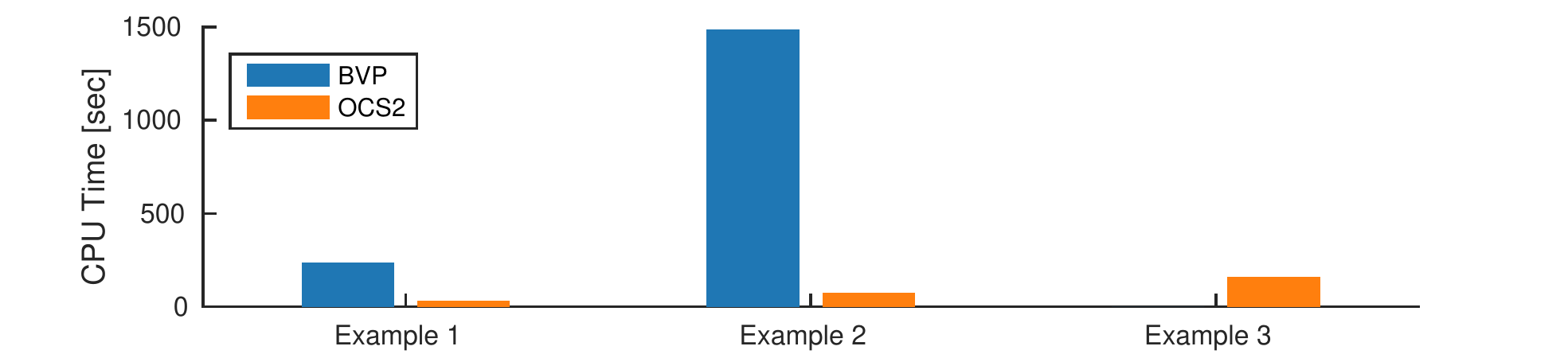}
    \caption{Comparison between the CPU time consumption of the BVP algorithm (blue) and the OCS2 algorithm (orange). The evaluation is done on an Intel's core i7 2.7-GHz processor. The BVP method run-time on Example 3 is missing since the algorithm failed to terminate.}
    \label{fig:cpu_time}
\end{figure}

\vspace{-3mm}
In this section, we evaluate the performance of the proposed method on three numerical examples. The first two examples are provided as illustrative cases to compare the performance of the proposed algorithm to the baseline algorithm introduced by \cite{xu04} and the third one is about the application of our method in motion control of legged robots. We here refer to the baseline algorithm as the BVP method since it is based on the solution of two-point BVPs.
\vspace{-4mm}
\paragraph*{Example 1:} 
The first example is a nonlinear switched system with three mode switches which are defined as followings
\begin{align*}
& 1: \left\{ \begin{array}{ll}
		\dot{x}_1 = x_1 + u_1\sin{x_1}\\
		\dot{x}_2 = -x_2 + u_1\cos{x_2}
		\end{array} 
\right.,   \quad
2: \left\{ \begin{array}{ll}
		\dot{x}_1 = x_2 + u_1\sin{x_2}\\
		\dot{x}_2 = -x_1 - u_1\cos{x_1}
		\end{array} 
\right.   \notag \\
& 3: \left\{ \begin{array}{ll}
		\dot{x}_1 = -x_1 - u_1\sin{x_1}\\
		\dot{x}_2 = x_2 + u_1\cos{x_2}
		\end{array} 
\right.
\end{align*}
with the initial condition $\vx_0 = \begin{bmatrix} 2, 3  \end{bmatrix}^\top$. The optimization goal is to calculate the optimal switching times $t_1$ (from subsystem 1 to 2), $t_2$ (from subsystem 2 to 3), and the continuous control input $u_1$ such that the following cost function is minimized
\begin{equation*} \label{eq:exp1_cost}
J = 0.5 \| \vx(3)-\vx_g \|^2 + 0.5 \int_0^3{ \left( \| \vx(t)-\vx_g \|^2 + \|\vu(t)\|^2 \right) dt } ,
\end{equation*} 
where $\vx_g = \begin{bmatrix} 1,& -1\end{bmatrix} ^\top$. We apply both OCS2 and BVP to this system with uniformly distributed initial switching times. The optimized switching times, the optimized cost, the number of the iterations, and number of function calls (bottom-level optimization calls) for both algorithms are presented in Table~\ref{tab:performance}. As it illustrates, OCS2 and BVP converge to the same solution within a comparable number of iterations and function calls. However, the consumed CPU times are significantly different. The BVP method utilizes about 235 seconds on an Intel's core i7 2.7-GHz processor while the OCS2 algorithm uses only 31 seconds which is roughly 7.5 times faster (Figure \ref{fig:cpu_time}).  
\vspace{-4mm}
\paragraph*{Example 2:} 
In order to examine the scalability of both algorithms to higher dimensions, we augment each subsystem in Example 1 with two more states and one additional control input. The augmented states have the following system dynamics
\begin{align*}
& 1': \left\{ \begin{array}{ll}
		\dot{x}_3 = -x_3 + 2 x_3 u_2 \\
		\dot{x}_4 = x_4 + x_4 u_2
		\end{array} 
\right.,   \quad
2': \left\{ \begin{array}{ll}
		\dot{x}_3 = x_3 - 3 x_3 u_2 \\
		\dot{x}_4 = 2 x_4 - 2 x_4 u_2
		\end{array} 
\right.   \notag \\
& 3': \left\{ \begin{array}{ll}
		\dot{x}_3 = 2 x_3 + x_3 u_2 \\
		\dot{x}_4 = -x_4 + 3 x_4 u_2
		\end{array} 
\right.
\end{align*}
with the initial condition $\vx_0 = \begin{bmatrix} 2, 3, 1, 1 \end{bmatrix}^{\top}$. The cost function is the same as Example 1 with $\vx_g = \begin{bmatrix} 1, -1, 2, 2\end{bmatrix} ^{\top}$. As Example 1, the optimized solutions are comparable for both algorithms (refer to Table~\ref{tab:performance}). However the CPU times are drastically different. For the BVP method, the CPU time is about 1500 seconds while the OCS2 algorithm uses only 76 seconds which is 19.5 times more efficient. This manifests the efficiency of the OCS2 algorithm in higher dimensional problems where computational time of the BVP algorithm prohibitively increases.

\begin{table}[t]
\centering 
\caption{Comparison between the performance of the BVP and OCS2 algorithms. In the table Itr. is the number of iterations in the gradient-descent method until it converges. FC. is the number of requests for cost function and its derivative evaluation in the gradient-descent algorithm.}
\label{tab:performance}
\vspace{-2mm}
\begin{tabular}{c c  c  c  c  c  c }
  \hline
  						& Alg.  & $J$    & $s_1$  & $s_2$  & Itr. & FC. \\
  \hline
  \multirow{2}{*}{EX1}  & BVP   & 5.4498 & 0.2235 & 1.0198 & 9    & 13 \\
      				    & OCS2  & 5.4438 & 0.2324 & 1.0236 & 7    & 14 \\
  \hline
  \hline
  \multirow{2}{*}{EX2}  & BVP   & 10.3797 & 0.2754 & 1.6076 & 9    & 13 \\
       					& OCS2  & 10.3888 & 0.2973 & 1.5978 & 8    & 20 \\
  \hline
\end{tabular}
\end{table}

\vspace{-4mm}
\paragraph*{Example 3: Quadruped's CoM motion control}
In this example, we apply OCS2 to the quadruped's CoM control problem. In a quadruped robot, based on the stance legs configuration the system shows different dynamical behavior. Furthermore, at the switching instances the non-elastic nature of contacts introduces a jump in the state trajectory. Therefore, basically a legged robot have to be modeled as a nonlinear hybrid system with discontinuous state trajectory.

The computational burden of solving the optimal control problem on such a nonlinear and hybrid system has encouraged researchers to use CoM dynamics instead of the complete system dynamics. In addition to optimizing a lower dimension problem, this approach has another important advantage: the CoM state trajectory is more smooth and the impact force does not cause a noticeable jump in states. Therefore the CoM dynamics can be modeled as a switched system which facilitates solving the optimal control problem. The CoM system has in total $12$ states consisting of: $3$ states for orientation, $\vth$, $3$ states for position, $\vp$, and $6$ states for linear and angular velocities in body frame, $\vv$ and $\vom$. The control inputs of the CoM system are an input which triggers the mode switch and the contact forces of the stance legs $\{\vlambda^i(t)\}_{i=1}^4$. The simplified CoM equation can be derived from the Newton-Euler equation as   
\begin{equation*}
\left\{ 
\begin{array}{ll}
		\dot{\vth} = \vR(\vth) \vom, \hspace{6mm}
		\dot{\vom} = \vI^{-1} \big( -\vom \times \vI\vom + \sum_{i=0}^4{\sigma_i \vJ_{\omega,i}^T \vlambda_i} \big), \\
		\dot{\vp}  = \vR(\vth) \vv, \hspace{8mm}
		\dot{\vv}  = \frac{1}{m} \big( m\vg + \sum_{i=0}^4{\sigma_i \vJ_{v,i}^T\vlambda_i} \big)  
\end{array}	 
\right.
\end{equation*}
where $\vR(\vth)$ is the rotation matrix, $\vg$ is gravitational acceleration in body frame, $\vI$ and $m$ are moment of inertia about the CoM and the total mass respectively. $\vJ_{\omega,i}$ and $\vJ_{v,i}$ are the Jacobian matrices of the $i$th foot with respect to CoM which depends on the foot position and CoM's position and orientation (for technical specifications of this quadruped robot refer to \cite{semini11}). In this example, we study the trotting gait. 

In a quadruped trotting gait, the pairwise alternating diagonal legs are the stance legs. In each phase of the motion only two opposing feet are on the ground. Since we assume that our robot has point feet, it looses controllability in one degree of freedom, namely the rotation around the connecting line between the two stance feet. Therefore, the subsystems in each phase of the trotting gait are not controllable individually. However, the ability to switch from one diagonal pair to the other, allows the robot for gaining control over that degree of freedom. Therefore, for a successful trot, the robot requires to plan and control the contact forces as well as the switching times.

The cost function in this example is defined as \eqref{eq:cost_quadratic_approximation}. The control effort weighting matrix is a diagonal matrix which penalizes all continuous control inputs (contact forces) equally. Furthermore, our cost function penalizes intermediate orientation and z-offsets during the entire trajectory. In the final cost, we penalize deviations from a target point $p = [3, 0, 0]^\top$ which lies 3~m in front of the robot. For the following evaluation, we assume that the x direction points to the front, the y direction points to the left and the z direction is orthogonal to ground. Furthermore, we penalize the linear and angular velocities to ensure the robot comes to a stop towards the end of the trajectory. 
 
\begin{figure}[t]
    \includegraphics[width=\columnwidth]{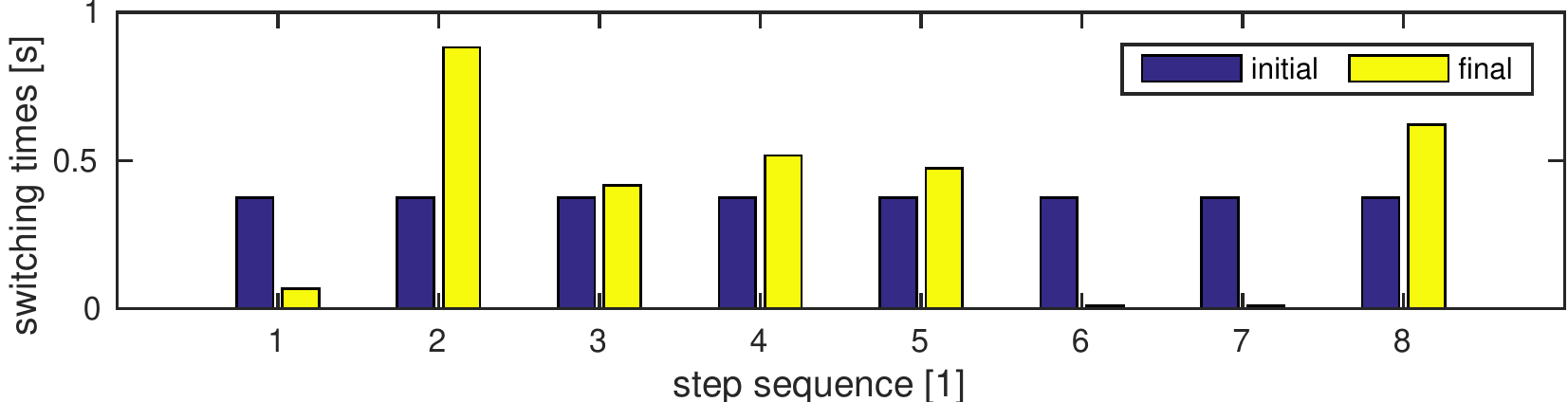}
    \caption{Comparison between the initial (blue) and optimized (yellow) switching times sequence. After optimization mode 6 and 7 switching times are set to zero and the number of total steps reduced to 6 form the initial 8 steps.}
    \label{fig:switching_times}
\end{figure}

Figure \ref{fig:switching_times} shows the difference between the initial and optimized switching times. The initial switching times have been chosen uniformly over the three-second period. During optimization the controller both extends and shortens different switching times.  More interestingly, it set the time periods of two phases to zero. Therefore, although we initially asked the robot to take 8 steps the optimized trot has 6 steps.

\begin{figure}[tbph]
    \includegraphics[width=\columnwidth]{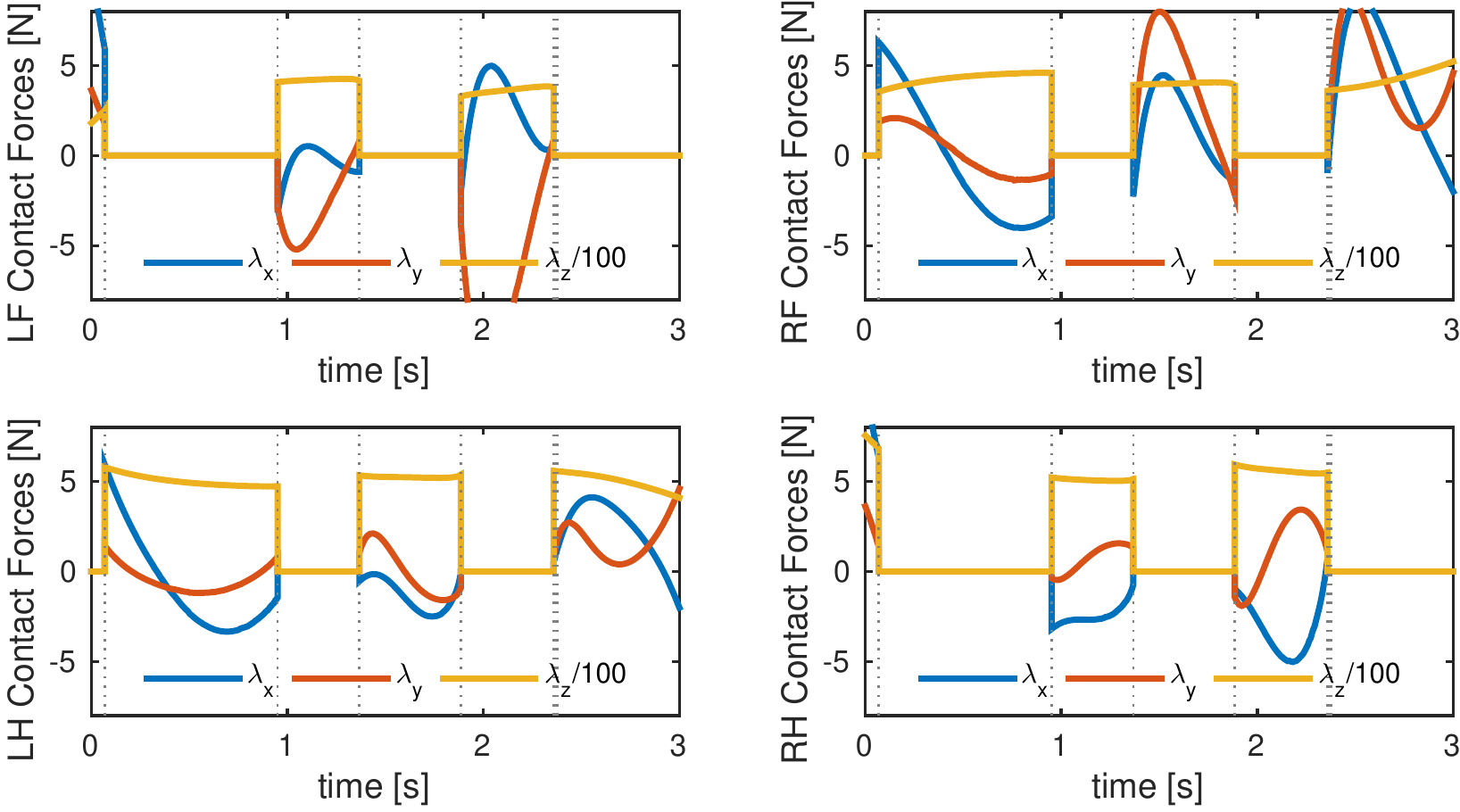}
    \caption{Contact forces as optimized by the trotting gait controller. For visibility, contact forces in z-direction have been scaled by a factor of 1/100. The gray dotted vertical lines indicate the switching times. The non-symmetric pattern results from the fact, that the quadruped is not perfectly symmetric and thus also not modeled as such.}
    \label{fig:contact_forces}
\end{figure}

Figure \ref{fig:contact_forces} shows the optimized contact forces of the trotting experiment. The vertical contact force $\lambda_z$ which needs to compensate for gravity is fairly equally distributed between the legs. However, we can see that the plots are not perfectly symmetric. This results from the quadruped being modeled after the hardware, which has unsymmetrical inertia. Additionally, we can see that the force profiles are non-trivial and thus would be difficult to derive manually. However, the contact forces are smooth within each step sequence which makes tracking easy.

\begin{figure}[tbph]
    \includegraphics[width=\columnwidth]{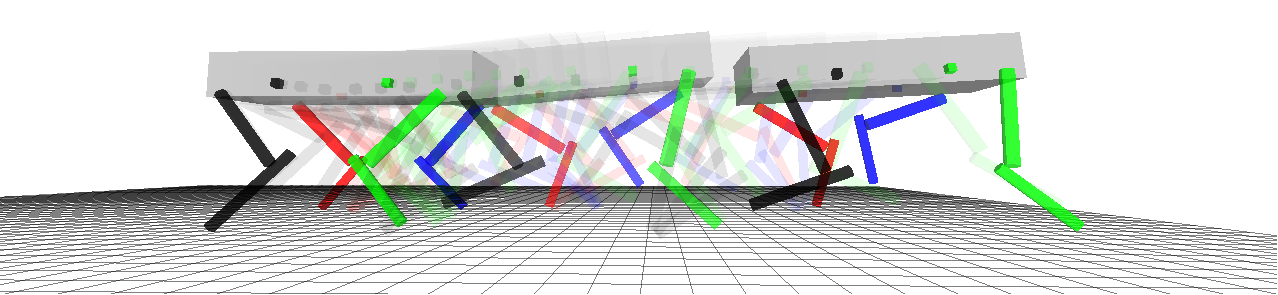}
    \caption{Few snapshots of the modeled quadruped during optimized trotting gait. The CoM motion is controlled by the controller synthesized by OCS2 algorithm. The swing foot trajectories (the ones that are not in contact) are designed based on a simple heuristic over the CoM linear velocity.}
    \label{fig:hyq}
\end{figure}

Figure \ref{fig:hyq} shows a few snapshots of the quadruped during the optimized trotting gait. The CoM motion is controlled by the OCS2-designed controller. The swing feet's trajectories (the ones that are not in contact) are designed based on a simple heuristic that the distance between the CoM and a swing foot at the moment of take off (breaking contact) and at the moment of touch down (establishing contact) should be the same.    

\vspace{-2mm}
\section{CONCLUSIONS AND FUTURE WORK}
\vspace{-3mm}
In this paper, we have presented an efficient method to solve the optimal control problem for nonlinear switched systems with predefined mode sequence. The proposed method is based on a two-stage optimization scheme which optimizes the continuous inputs as well as the switching times. In order to obtain an accurate estimation of the cost function derivatives, \cite{xu04} have introduced a set of two-point BVPs. Although there exist many numerical methods for solving two-point BVP (e.g. collocation method), many of them do not scale properly to high dimensional problems such as controlling quadruped locomotion. To tackle this issue, we have proposed the OCS2 algorithm which is based on an SLQ algorithm. In order to obtain cost function derivatives, OCS2 takes into account the sensitivity of the approximated LQ models with respect to the switching times. OCS2 obtains an approximation of the LQ model sensitivity, only by using the LQ approximation of the problem which has been already calculated by SLQ. Therefore, the algorithm can calculate values of the derivatives with no further computational cost for evaluating the higher order derivatives of the system dynamics and cost function. This feature increases the proposed algorithm efficiency in comparison to other methods such as direct collocation which rely on higher order derivatives. 

In order to demonstrate the computational efficiency of the algorithm, we have compared the CPU time used by the OCS2 algorithm with the baseline method introduced by \cite{xu04}. We observe that as the dimensions of state and input spaces increase, the difference in computational time between the algorithms becomes significant to the point that for the quadruped CoM control task, the run-time of the baseline algorithm becomes prohibitively long.    




\vspace{-2mm}
\bibliography{bibliography/references}

\end{document}

%% file: mathdef.tex
\newcommand{\argmin}{\mathop{\mathrm{argmin}}}

\newcommand{\vth}{\mbox{\boldmath $\theta$}}

\newcommand{\vlambda}{\mbox{\boldmath $\lambda$}}

\newcommand{\vom}{\mbox{\boldmath $\omega$}}

\newcommand{\vf}{\mathbf f}
\newcommand{\vg}{\mathbf g}

\newcommand{\vl}{\mathbf l}

\newcommand{\vp}{\mathbf p}
\newcommand{\vq}{\mathbf q}
\newcommand{\vr}{\mathbf r}
\newcommand{\vs}{\mathbf s}
\newcommand{\vt}{\mathbf t}
\newcommand{\vu}{\mathbf u}
\newcommand{\vv}{\mathbf v}
\newcommand{\vw}{\mathbf w}
\newcommand{\vx}{\mathbf x}
\newcommand{\vy}{\mathbf y}

\newcommand{\vA}{\mathbf A}
\newcommand{\vB}{\mathbf B}

\newcommand{\vI}{\mathbf I}
\newcommand{\vJ}{\mathbf J}

\newcommand{\vL}{\mathbf L}

\newcommand{\vP}{\mathbf P}
\newcommand{\vQ}{\mathbf Q}
\newcommand{\vR}{\mathbf R}
\newcommand{\vS}{\mathbf S}

\newcommand{\vW}{\mathbf W}





%% file: root.bbl
\begin{thebibliography}{25}
\providecommand{\natexlab}[1]{#1}
\providecommand{\url}[1]{\texttt{#1}}
\providecommand{\urlprefix}{URL }
\expandafter\ifx\csname urlstyle\endcsname\relax
  \providecommand{\doi}[1]{doi:\discretionary{}{}{}#1}\else
  \providecommand{\doi}{doi:\discretionary{}{}{}\begingroup
  \urlstyle{rm}\Url}\fi

\bibitem[{Armijo(1966)}]{armijo96}
Armijo, L. (1966).
\newblock Minimization of functions having lipschitz continuous first partial
  derivatives.
\newblock \emph{Pacific Journal of mathematics}.

\bibitem[{Bemporad and Morari(1999)}]{bemporad99}
Bemporad, A. and Morari, M. (1999).
\newblock Control of systems integrating logic, dynamics, and constraints.
\newblock \emph{Automatica}.

\bibitem[{Bengea and DeCarlo(2005)}]{bengea05}
Bengea, S. and DeCarlo, R. (2005).
\newblock Optimal control of switching systems.
\newblock \emph{Automatica}.

\bibitem[{Borrelli et~al.(2005)Borrelli, Baoti{\'c}, Bemporad, and
  Morari}]{borrelli05}
Borrelli, F., Baoti{\'c}, M., Bemporad, A., and Morari, M. (2005).
\newblock Dynamic programming for constrained optimal control of discrete-time
  linear hybrid systems.
\newblock \emph{Automatica}.

\bibitem[{Branicky et~al.(1998)Branicky, Borkar, and Mitter}]{branicky98}
Branicky, M.S., Borkar, V.S., and Mitter, S.K. (1998).
\newblock A unified framework for hybrid control: model and optimal control
  theory.
\newblock \emph{IEEE Transactions on Automatic Control}.

\bibitem[{Bryson(1975)}]{bryson75}
Bryson, A.E. (1975).
\newblock \emph{Applied optimal control: optimization, estimation and control}.
\newblock CRC Press.

\bibitem[{Egerstedt et~al.(2003)Egerstedt, Wardi, and Delmotte}]{egerstedt03}
Egerstedt, M., Wardi, Y., and Delmotte, F. (2003).
\newblock Optimal control of switching times in switched dynamical systems.
\newblock In \emph{42nd IEEE Conference on Decision and Control (CDC)}.

\bibitem[{Giua et~al.(2001)Giua, Seatzu, and der Mee}]{giua01}
Giua, A., Seatzu, C., and der Mee, C.V. (2001).
\newblock Optimal control of switched autonomous linear systems.
\newblock In \emph{40th IEEE Conference on Decision and Control (CDC)}.

\bibitem[{Jaggi(2013)}]{jaggi13}
Jaggi, M. (2013).
\newblock Revisiting frank-wolfe: Projection-free sparse convex optimization.
\newblock In \emph{ICML}.

\bibitem[{Johnson and Murphey(2011)}]{johnson11}
Johnson, E.R. and Murphey, T.D. (2011).
\newblock Second-order switching time optimization for nonlinear time-varying
  dynamic systems.
\newblock \emph{IEEE Transactions on Automatic Control}.

\bibitem[{Kamgarpour and Tomlin(2012)}]{kamgarpour12}
Kamgarpour, M. and Tomlin, C. (2012).
\newblock On optimal control of non-autonomous switched systems with a fixed
  mode sequence.
\newblock \emph{Automatica}.

\bibitem[{Mayne(1966)}]{mayne66}
Mayne, D. (1966).
\newblock A second-order gradient method for determining optimal trajectories
  of non-linear discrete-time systems.
\newblock \emph{International Journal of Control}.

\bibitem[{Neunert et~al.(2016)Neunert, de~Crousaz, Furrer, Kamel, Farshidian,
  Siegwart, and Buchli}]{neunert16}
Neunert, M., de~Crousaz, C., Furrer, F., Kamel, M., Farshidian, F., Siegwart,
  R., and Buchli, J. (2016).
\newblock Fast nonlinear model predictive control for unified trajectory
  optimization and tracking.
\newblock In \emph{IEEE International Conference on Robotics and Automation
  (ICRA)}.

\bibitem[{Pakniyat and Caines(2014)}]{pakniyat14}
Pakniyat, A. and Caines, P.E. (2014).
\newblock On the relation between the minimum principle and dynamic programming
  for hybrid systems.
\newblock In \emph{53rd IEEE Conference on Decision and Control}.

\bibitem[{Riedinger et~al.(2003)Riedinger, Iung, and Kratz}]{riedinger03}
Riedinger, P., Iung, C., and Kratz, F. (2003).
\newblock An optimal control approach for hybrid systems.
\newblock \emph{European Journal of Control}.

\bibitem[{Riedinger et~al.(1999)Riedinger, Kratz, Iung, and
  Zanne}]{riedinger99}
Riedinger, P., Kratz, F., Iung, C., and Zanne, C. (1999).
\newblock Linear quadratic optimization for hybrid systems.
\newblock In \emph{38th IEEE Conference on Decision and Control (CDC)}.

\bibitem[{Semini et~al.(2011)Semini, Tsagarakis, Guglielmino, Focchi, Cannella,
  and Caldwell}]{semini11}
Semini, C., Tsagarakis, N.G., Guglielmino, E., Focchi, M., Cannella, F., and
  Caldwell, D.G. (2011).
\newblock Design of hyq--a hydraulically and electrically actuated quadruped
  robot.
\newblock \emph{Proceedings of the Institution of Mechanical Engineers, Part I:
  Journal of Systems and Control Engineering}.

\bibitem[{Shaikh and Caines(2007)}]{shaikh07}
Shaikh, M.S. and Caines, P.E. (2007).
\newblock On the hybrid optimal control problem: theory and algorithms.
\newblock \emph{IEEE Transactions on Automatic Control}.

\bibitem[{Sideris and Bobrow(2005)}]{sideris05}
Sideris, A. and Bobrow, J. (2005).
\newblock An efficient sequential linear quadratic algorithm for solving
  nonlinear optimal control problems.
\newblock \emph{IEEE Transactions on Automatic Control}.

\bibitem[{Sideris and Rodriguez(2010)}]{sideris10}
Sideris, A. and Rodriguez, L.A. (2010).
\newblock A riccati approach to equality constrained linear quadratic optimal
  control.
\newblock In \emph{Proceedings of the 2010 American Control Conference}.

\bibitem[{Soler et~al.(2012)Soler, Kamgarpour, Tomlin, and Staffetti}]{soler12}
Soler, M., Kamgarpour, M., Tomlin, C., and Staffetti, E. (2012).
\newblock Multiphase mixed-integer optimal control framework for aircraft
  conflict avoidance.
\newblock In \emph{51st IEEE Conference on Decision and Control (CDC)}.

\bibitem[{Sussmann(1999)}]{sussmann99}
Sussmann, H.J. (1999).
\newblock A maximum principle for hybrid optimal control problems.
\newblock In \emph{38th IEEE Conference on Decision and Control (CDC)}.

\bibitem[{Todorov and Li(2005)}]{todorov05}
Todorov, E. and Li, W. (2005).
\newblock A generalized iterative lqg method for locally-optimal feedback
  control of constrained nonlinear stochastic systems.
\newblock In \emph{Proceedings of the 2005 American Control Conference}.

\bibitem[{Wardi et~al.(2012)Wardi, Egerstedt, and Twu}]{wardi12}
Wardi, Y., Egerstedt, M., and Twu, P. (2012).
\newblock A controlled-precision algorithm for mode-switching optimization.
\newblock In \emph{51st IEEE Conference on Decision and Control (CDC)}.

\bibitem[{Xu and Antsaklis(2004)}]{xu04}
Xu, X. and Antsaklis, P.J. (2004).
\newblock Optimal control of switched systems based on parameterization of the
  switching instants.
\newblock \emph{IEEE Transactions on Automatic Control}.

\end{thebibliography}
